\documentclass[11pt,titlepage,a4paper]{article}
\usepackage{bozhomac}	
\usepackage{bozhlogo}

\title{\bfseries	\vspace*{-2.1095in}
\vspace*{-3ex}
{
\begin{flushright}
	\textbf{\large LANL xxx E-print archive No. quant-ph/9901039}\\[2ex]
\end{flushright}
}
{\huge Fibre bundle formulation of	\\[6pt]
nonrelativistic quantum mechanics	\\[2ex]
\hspace*{-1.123456ex}\mbox{\Large
		IV. Mixed states and evolution transport's curvature}
}
}

\vspace{3ex}

\author{
Bozhidar Z. Iliev
\thanks{Department Mathematical Modeling,
Institute for Nuclear Research and \mbox{Nuclear} Energy,
Bulgarian Academy of Sciences,
Boul. Tzarigradsko chauss\'ee~72, 1784 Sofia, Bulgaria}
\thanks{E-mail address: bozho@inrne.bas.bg}
\thanks{URL: http://www.inrne.bas.bg/mathmod/bozhome/}
}

\date{
\vspace{2.27ex}\ShortTitle{Bundle quantum mechanics: IV}	\\[0.27ex]
\vspace{3.27ex}
	\begin{tabular}{r@{$\colon\to~$}l}
\vspace{0.09ex} Basic ideas	& March 1996		\\[0.09ex]
\vspace{0.09ex} Began		& May 19, 1996		\\[0.09ex]
\vspace{0.09ex} Ended		& July 12, 1996		\\[0.09ex]
\vspace{0.09ex} Revised		& December 1996 -- January 1997,\\[0.09ex]
\vspace{0.09ex} Revised		& April 1997, September 1998	\\[0.09ex]
\vspace{0.09ex} Last update	& October 20, 1998  	\\[0.09ex]
\vspace{0.09ex} Composing/Extracting part IV
			&  September 27/October 4, 1997	\\[0.09ex]
\vspace{0.09ex} Updating part IV   &  February 14, 1999	\\[0.09ex]
\vspace{0.27ex} Produced	& \fbox{\today}		\\[0.27ex]
	\end{tabular} \\[1.27ex]
	\begin{tabular}{r@{$\colon~$}l}
\vspace{0.27ex} LANL xxx archive server E-print No.& quant-ph/9901039
							\\[0.27ex]
	\end{tabular} \\[-0.27ex]
\vspace{3.27ex}{\Huge\BOZHO}	\\[3.27ex]
\vspace{0.27ex}\Subject{Quantum mechanics; Differential geometry} \\[2.27ex]
	\begin{tabular}{r@{\hspace{0.512em}}|@{\hspace{0.512em}}l}
\vspace{0.27ex}		  
&
\vspace{0.27ex}
	\end{tabular} \\[1.27ex]
\vspace{0.27ex}	\\[0.27ex]
}
\newcommand{\Hil}{\mathcal{F}}	
\newcommand{\HilB}{(\bHil,\proj,\base)}	
	\newcommand{\bHil}{\mathit{F}}	
	\newcommand{\proj}{\pi}		
	\newcommand{\base}{\mathit{M}}	

\newcommand{\Ham}{\mathcal{H}}	
\newcommand{\bHam}{\mathit{H}}	

\newcommand{\HamM}{\boldsymbol{\Ham}} 
\newcommand{\bHamM}{\boldsymbol{\bHam}} 

\newcommand{\mbHam}{\boldsymbol{\bHam}\!^\mathbf{m}}

\newcommand{\dyn}[1]{\pmb{\mathbb{#1}}}	
	\newcommand{\ope}[1]{\mathcal{#1}}		 
	\newcommand{\mor}[1]{\mathit{#1}}		 
	\newcommand{\mmor}[1]{\boldsymbol{\mathit{#1}}}	 

\newcommand{\ih}{\mathrm{i}\hbar}
\newcommand{\iih}{\frac{1}{\ih}} 

\newcommand{\Rho}{\mathrm{P}} 

\begin{document}		


\renewcommand*{\thesection}{I.\arabic{section}}
\renewcommand*{\thefootnote}{\protect{I.\arabic{footnote}}}
\setcounter{page}{0}
\setcounter{footnote}{0}
\setcounter{section}{0}
\setcounter{equation}{0}
 	\include{bqm-1txt}
\renewcommand*{\thesection}{\arabic{section}}
\renewcommand*{\thefootnote}{\arabic{footnote}}
\setcounter{page}{0}
\setcounter{footnote}{0}
\setcounter{section}{0}
\setcounter{equation}{0}

\renewcommand*{\thesection}{II.\arabic{section}}
\renewcommand*{\thefootnote}{\protect{II.\arabic{footnote}}}
\setcounter{page}{0}
\setcounter{footnote}{0}
\setcounter{section}{0}
\setcounter{equation}{0}
 	\include{bqm-2txt}
\renewcommand*{\thesection}{\arabic{section}}
\renewcommand*{\thefootnote}{\arabic{footnote}}
\setcounter{page}{0}
\setcounter{footnote}{0}
\setcounter{section}{0}
\setcounter{equation}{0}

\renewcommand*{\thesection}{III.\arabic{section}}
\renewcommand*{\thefootnote}{\protect{III.\arabic{footnote}}}
\setcounter{page}{0}
\setcounter{footnote}{0}
\setcounter{section}{0}
\setcounter{equation}{0}
 	\include{bqm-3txt}
\renewcommand*{\thesection}{\arabic{section}}
\renewcommand*{\thefootnote}{\arabic{footnote}}
\setcounter{page}{0}
\setcounter{footnote}{0}
\setcounter{section}{0}
\setcounter{equation}{0}


\renewcommand{\thefootnote}{\fnsymbol{footnote}}
\maketitle			
\renewcommand{\thefootnote}{\arabic{footnote}}

\tableofcontents		


\pagestyle{myheadings}
\markright{\itshape\bfseries Bozhidar Z. Iliev:
	\upshape\sffamily\bfseries Bundle quantum mechanics.~IV}

\begin{abstract}

We propose a new systematic fibre bundle formulation of nonrelativistic
quantum mechanics. The new form of the theory is equivalent to the usual one
but it is in harmony with the modern trends in theoretical physics and
potentially admits new generalizations in different directions. In it
a pure state of some quantum system is described by a state section (along
paths) of a (Hilbert) fibre bundle. It's evolution is determined through the
bundle (analogue of the) Schr\"odinger equation. Now the dynamical variables
and the density operator are described via bundle morphisms (along paths).
The mentioned quantities are connected by a number of relations derived in
this work.
%
%

	The present fourth part of this series is devoted mainly to the fibre
bundle description of mixed quantum states. We show that to the conventional
density operator there corresponds a unique density morphism (along paths)
for which the corresponding equations of motion are derived. It is also
investigated the bundle description of mixed quantum states in the different
pictures of motion. We calculate the curvature of the evolution transport and
prove that it is curvature free iff the values of the Hamiltonian operator at
different moments commute.

\end{abstract}

\section {Introduction}
\label{introduction-IV}
\setcounter{equation} {0}

	This is the fourth part of our series of works considering the
application of the theory of fibre bundles to nonrelativistic quantum
mechanics.

	The paper, which is a straightforward continuation
of~\cite{bp-BQM-introduction+transport,bp-BQM-equations+observables,
						bp-BQM-pictures+integrals},
is organized as follows.

	The bundle approach to the quantum mechanics of mixed states is
investigated in Sect.~\ref{XI}. Subsection~\ref{XI.1} is a brief review of
the conventional concepts of mixed state(s) and density operator (matrix).
Their bundle description is presented in Subsection~\ref{XI.2}. It turns out
that to the density operator there corresponds a suitable
\emph{density morphism along paths}. The equations for its time evolution
are derived. In Subsection~\ref{XI.3} are studied problems connected with the
representations and description of mixed quantum states in the different
pictures of motion. The equations of motion for density morphisms and
operators are derived.

	Sect.~\ref{IX} is devoted to the curvature of the (bundle) evolution
transport.

	The work ends with some remarks in Sect.~\ref{conclusion-IV}.
\vspace{1.2ex}

	The notation of the present part of the series is identical with
the one of the preceding parts for which the reader is referred
to~\cite{bp-BQM-introduction+transport,bp-BQM-equations+observables,
						bp-BQM-pictures+integrals}.

	The references to sections, equations, footnotes etc.
from the previous three parts of the series,
namely~\cite{bp-BQM-introduction+transport}, \cite{bp-BQM-equations+observables}
			and~\cite{bp-BQM-pictures+integrals},
are denoted by the corresponding sequential reference numbers in these parts
preceded by the Roman number of the part in which it appears and a dot as a
separator. For instance, Sect.~I.5 and (III.3.16) mean respectively
section 5 of part~I, i.e.\  of~\cite{bp-BQM-introduction+transport}, and
equation~(3.16) (equation~16 in Sect.~3) of
part~III, i.e.\  of~\cite{bp-BQM-pictures+integrals}.

	Below, for reference purposes, we present a list of some essential
equations
of~\cite{bp-BQM-introduction+transport,bp-BQM-equations+observables,
					bp-BQM-pictures+integrals}
used in this paper.  Following the above convention, we retain their
original reference numbers.
	\enlargethispage*{2ex}
	\begin{gather}
\tag{\ref{4.3}}
\psi(t) = l_{\gamma(t)} ( {\Psi}_\gamma(t) ) \in \Hil ,
\\
\tag{\ref{4.10b}}
l_{x}^{\ddag} = l_{x}^{-1},
\\
\tag{\ref{4.7}}
\mor{U}_\gamma(t,s) =
		l_{\gamma(t)}^{-1}\circ \ope{U}(t,s) \circ l_{\gamma(s)},
\qquad s,t\in J ,
\\
\tag{\ref{5.10}}
\boldsymbol{\Gamma}_\gamma(t) :=
\left[ \Gamma_{{\ }a}^{b}(t;\gamma) \right] =
- \iih \mbHam_{\gamma}(t) ,
\\
\tag{\ref{6.0}}
\mor{A}_{\gamma}(t) =
l_{\gamma(t)}^{-1} \circ \ope{A}(t) \circ l_{\gamma(t)}
\colon  \bHil_{\gamma(t)}\to\bHil_{\gamma(t)} ,
\\
\tag{\ref{6.5}}
\pmb{\boldsymbol{[}} \tilde{\mor{D}}_{t}^{\gamma} (C) \pmb{\boldsymbol{]}} =
\frac{d}{dt} \boldsymbol{C}_t +
\left[
\boldsymbol{\Gamma}_\gamma(t),\boldsymbol{C}_t
\right]_{\_} ,
\\
\tag{\ref{7.4}}
\mor{A}_{\gamma,t}^{\mathrm{H}}(t_0) :=
\mor{U}_\gamma^{-1}(t,t_0)\circ \mor{A}_\gamma(t)\circ \mor{U}_\gamma(t,t_0)
\colon  \bHil_{\gamma(t_0)}\to  \bHil_{\gamma(t_0)} ,
\\
\tag{\ref{7.7}}
\ope{A}_{t}^{\mathrm{H}}(t_0) :=
\ope{U}(t_0,t)\circ\ope{A}(t)\circ\ope{U}(t,t_0)\colon \Hil\to\Hil ,
\\
\tag{\ref{7.10}}
\ih\frac{\partial \mor{A}_{\gamma,t}^{\mathrm{H}}(t_0)}{\partial t} =
\left[
\mor{A}_{\gamma,t}^{\mathrm{H}}(t_0), H_{\gamma,t}^{\mathrm{H}}(t_0)
\right] _{\_} +
\ih\left(
\frac{\partial\ope{A}}{\partial t}
\right)_{\!\gamma,t}^{\!\mathrm{H}} (t_0) ,
\\
\tag{\ref{7.16}}
\mor{A}_{\gamma,t}^\mor{V}(t_1) :=
\mor{V}_\gamma(t_1,t) \circ \mor{A}_\gamma(t) \circ
\mor{V}_\gamma^{-1}(t_1,t)\colon
\bHil_{\gamma(t_1)} \to \bHil_{\gamma(t_1)} ,
	\end{gather}
	\begin{gather}
\tag{\ref{7.27}}
\ih \frac{\partial\ope{A}_{t}^{\ope{V}}(t_1)}{\partial t} =
\left[
\ope{A}_{t}^{\ope{V}}(t_1), _\ope{V}\!\Ham_{t}^{\ope{V}}(t_1)
\right]_{\_}
+
\ih\left( \frac{\partial\ope{A}(t)}{\partial t} \right)
_{\!t}^{\!\ope{V}} (t_1) ,
\\
\tag{\ref{7.29}}
\ih \frac{\partial \mor{A}_{\gamma,t}^{\mor{V}}(t_1) }{\partial t}     =
\left[
\mor{A}_{\gamma,t}^{\mor{V}}(t_1), _\mor{V}\!\bHam_{\gamma,t}^{\mor{V}}(t_1)
\right]_{\_}
+
\ih
\left(
\frac{\partial \ope{A}(t)} {\partial t}
\right)_{\!\gamma,t}^{\!\mor{V}}(t_1) .
	\end{gather}

\section{Mixed states}
\label{XI}
\setcounter{equation}{0}

	In the framework of quantum mechanics the most general description of
a state of a quantum system is provide via the so-called density (or
statistical) matrix (or operator) by means of which is achieved a uniform
description of pure and mixed states
(see~\cite[chapter~VIII, sect.~IV]{Messiah-QM} and~\cite[\S~33]{Dirac-PQM};
for mathematically rigorous exposition of the problem
see~\cite[chapter~IV, sect.~8]{Prugovecki-QMinHS}). This formalism has also
a bundle analogue which is described in this section.

\subsection{Hilbert space description (review)}
\label{XI.1}

	Here we briefly recall the notions of a mixed state and density
operator in the conventional Hilbert space description of quantum
mechanics~\cite{Messiah-QM} (see also~\cite{Uhlmann-91a, Anandan-90a}).

	Consider a quantum system which at a moment
$t$ with a probability $p_i$ can be found in a state with a state vector
$\psi_i(t)\in\Hil$. Here $i$ belongs to some set of indexes $I$ and the
statistical weights $p_i$ are assumed time-independent:
	\begin{equation}	\label{11.1}
0\le p_i \le 1,\quad \sum_{i\in I}p_i = 1,\quad
\frac{\partial p_i}{\partial t} = 0.
	\end{equation}
The state of such a system is described by the \emph{density operator}
	\begin{equation}	\label{11.2}
\rho(t) := \sum_{i\in I}\psi_i(t)
\frac{p_i}{\langle\psi_i(t)|\psi_i(t)\rangle}
\psi_i^{\dagger}(t)
\colon\Hil\to\Hil.
	\end{equation}
Here with a dagger as a superscript we denote the dual Hermitian conjugate
vectors and spaces with respect to the inner product
$\langle\cdot|\cdot\rangle$, i.e.\  if $\psi\in\Hil$, then
$\psi^\dagger\in\Hil^\dagger$ is a map
$\psi^\dagger\colon\Hil\to\mathbb{C}$ such that
 $\psi^\dagger\colon\chi\mapsto\langle\psi|\chi\rangle$ for
 $\chi\in\Hil$, and a product like
 $\psi\chi^\dagger$,  $\psi,\chi\in\Hil$ is defined as an operator
 $\psi\chi^\dagger\colon\Hil\to\Hil$ via
 $(\psi\chi^\dagger)(\varphi):=(\chi^\dagger(\varphi))\psi
			      =\langle\chi|\varphi\rangle\psi$ for
 $\varphi\in\Hil$.%
\footnote{%
In Dirac's notation  $\psi,\ \psi^\dagger,$ and $\psi\chi^\dagger$ will look
like $|\psi\rangle,\ \langle\psi|$ and $|\psi\rangle\langle\chi|$
respectively~\cite{Dirac-PQM,Messiah-QM}. Notice that
 $(\psi\chi^\dagger)^\dagger=\chi\psi^\dagger$ corresponds to
 $( |\psi\rangle\langle\chi| )^\dagger = |\chi\rangle\langle\psi|$
in Dirac's notation.%
}
The density operator~\eqref{11.2} is Hermitian, positive definite,
trace-class, and of unit trace~\cite{Messiah-QM,Prugovecki-QMinHS}. So we
have
	\begin{equation}	\label{11.3}
\rho^\dagger(t) = \rho(t),\quad
\langle\psi(t)|\rho(t)\psi(t)\rangle \ge 0,\quad
\Tr\rho(t) = 1,
	\end{equation}
where $\Tr$ denotes the trace of an operator. Conversely, any such operator
is a density operator (in the absence of superselection
rules)~\cite[chapter~IV, subsect.~8.6]{Prugovecki-QMinHS}.

	By definition the mean (expectation) value of an observable $\ope{A}$
is
	\begin{equation}	\label{11.4}
\langle\ope{A}(t)\rangle_{\rho}^{t} :=
	\Tr( \rho(t)\circ\ope{A}(t) )
	\end{equation}
for a system whose state is described by a density operator $\rho(t)$.

	The time evolution of the density operator is described by
postulating the Schr\"odinger equation for it, called also
\emph{von Neumann's or Liouville equation}:
	\begin{equation}	\label{11.5}
\ih \frac{d\rho(t)}{dt} =
[ \Ham(t),\rho(t) ]_{\_}  := \Ham(t)\circ\rho(t) - \rho(t)\circ\Ham(t)
	\end{equation}
where $\Ham(t)$ is system's Hamiltonian. If $\rho(t_0)$ is known for some
instant of time $t_0$, then
	\begin{equation}	\label{11.6}
\rho(t) = \ope{U}(t,t_0)\circ\rho(t_0)\circ\ope{U}^{-1}(t,t_0),
	\end{equation}
where $\ope{U}(t,t_0)$ is the system's evolution operator (see
Sect.~\ref{II}). In fact,~\eqref{11.6} is the general solution of~(\ref{11.5})
with respect to $\rho(t)$.%
\footnote{%
Equation~(\ref{11.5}) is equivalent to the assumption that every vector
$\psi_i(t)$ in~\eqref{11.2} evolves according to the Schr\"odinger
equation~(\ref{2.5}). Respectively, equation~(\ref{11.6}) is
equivalent to ~\eqref{2.1} for the vectors $\psi_i(t).$%
}

	If the sum in~\eqref{11.2} contains only one term or terms which are
proportional up to a phase factor to one of them, the system is said to be in
a \emph{pure state} (described by (one of) the corresponding state vector(s)
in~\eqref{11.2}), otherwise the system's state is called \emph{mixed}. A
criterion for a state to be pure is
$\rho^2(t)=\rho(t)$~\cite{Messiah-QM,Prugovecki-QMinHS}.

\subsection{Hilbert bundle description}
\label{XI.2}

	Now we are ready to apply the bundle approach to systems described
via some density operator $\rho(t)$. The easiest way to introduce the bundle
analogue of $\rho(t)$ is via~(\ref{11.4}). Expressing $\ope{A}$
from~\eqref{6.0} and substituting the result into~\eqref{11.4}, we find
	\begin{equation}	\label{11.7}
\langle \ope{A}(t) \rangle_{\rho}^{t} =
\langle \mor{A}_\gamma(t) \rangle_{\Rho_\gamma}^{t}
	\end{equation}
where
	\begin{equation}	\label{11.8}
\langle \mor{A}_\gamma(t) \rangle_{\Rho_\gamma}^{t} =
\Tr( \Rho_\gamma(t)\circ\mor{A}_\gamma(t) )
	\end{equation}
is the (bundle) mean value of the morphism $\mor{A}$, corresponding to
$\ope{A}$, in the state characterized by the
\emph{density morphism along paths}
 \(
\Rho\colon\gamma\mapsto\Rho_\gamma
\colon\gamma(t)\mapsto \{ \Rho_\gamma(s) : s\in J,\ \gamma(s)=\gamma(t) \}
 \)
defined via (cf.~\eqref{6.0})
	\begin{equation}	\label{11.9}
\Rho_\gamma(t) := l_{\gamma(t)}^{-1}\circ\rho(t)\circ l_{\gamma(t)}^{}
\colon\bHil_{\gamma(t)}\to\bHil_{\gamma(t)}.
	\end{equation}

	Meanwhile, equation~(\ref{11.7}) expresses the natural requirement
that the expectation value of a dynamical variable $\dyn{A}$ must
be independent of the (mathematical) way we calculate it.

	The bundle density operator has also a representation
like~(\ref{11.2}). In fact, substituting~\eqref{4.3} and its Hermitian
conjugate, i.e
	\begin{equation}	\label{11.10}
\psi^\dagger(t) = \Psi_\gamma^\ddagger(t)\circ l_{\gamma(t)},
	\end{equation}
where
 $\Psi_x^\ddagger\colon\bHil_x\to\mathbb{C}$ is defined by
\(
\Psi_x^\ddagger\colon\Phi_x\mapsto \langle \Psi_x | \Phi_x \rangle_x
\)
for $\Psi_x,\Phi_x\in\bHil_x$, $x\in \base$,
which is a consequence of the unitarity of $l_x$ (see~\eqref{4.10b}), for the
vectors $\psi_i(t)$ (appearing in~\eqref{11.2}) into~\eqref{11.2}, we get
	\begin{equation}	\label{11.11}
\rho(t) = l_{\gamma(t)}^{}\circ\Rho_\gamma(t)\circ l_{\gamma(t)}^{-1}
	\end{equation}
which is equivalent to~\eqref{11.9} with
	\begin{equation}	\label{11.12}
\Rho_\gamma(t) =
\sum_{i\in I} \Psi_{i,\gamma}(t)
\frac{p_i}{\langle\Psi_{i,\gamma}(t) | \Psi_{i,\gamma}(t)\rangle}
\Psi_{i,\gamma}^\ddagger(t).
	\end{equation}
where a product like $\Phi_x\Psi_x^\ddagger$ is considered as an operator
$\Phi_x\Psi_x^\ddagger\colon\bHil_x\to\bHil_x$ such that
 \(
(\Phi_x\Psi_x^\ddagger)\mathrm{X}_x :=
\langle\Psi_x | \mathrm{X}_x\rangle_x \Phi_x,\
\mathrm{X}_x\in\bHil_x.
 \)

	The above results show that the transition from Hilbert space to
Hilbert bundle description of mixed states is achieved simply via a
replacement of the vectors (resp.\ operators) of (resp.\ acting on) $\Hil$ with
sections (resp.\ morphisms) along paths of $\HilB$ according to the general
rules of sections~\ref{new-I} and~\ref{VI}. As we shall see below, this
observation has a general validity.

	From~\eqref{11.3}, \eqref{11.9} and~\eqref{4.10b} follows that the
density operator $\Rho_\gamma(t)$  in $\bHil_{\gamma(t)}$ is Hermitian,
positive definite, trace-class, and of trace one, i.e.\
	\begin{equation}	\label{11.13}
\Rho_\gamma^\ddagger(t) = \Rho_\gamma(t),\quad
	\langle
	\Psi_\gamma(t) | \Rho_\gamma(t)\Psi_\gamma(t)
	\rangle_{\gamma(t)} \ge 0,\quad
\Tr\left( \Rho_\gamma(t) \right) = 1.
	\end{equation}

	The time evolution of the density morphism along paths, i.e.\  the
relation between $\Rho_\gamma(t)$ and $\Rho_\gamma(t_0)$ for any
$t,t_0\in J$, can be found as follows. Substituting~\eqref{11.11} for $t=t_0$
into~\eqref{11.6}, then substituting the result into~\eqref{11.9}, and,
at last, applying~\eqref{4.7}, we get
	\begin{equation}	\label{11.14}
\Rho_\gamma(t) =
\mor{U}_\gamma(t,t_0)\circ\Rho_\gamma(t_0)\circ\mor{U}_{\gamma}^{-1}(t,t_0),
\qquad t,t_0\in J,
	\end{equation}
where $\mor{U}_\gamma(t,t_0)$ is the evolution transport from
$\bHil_{\gamma(t_0)}$ to $\bHil_{\gamma(t)}$.

	The differential equation, corresponding to the evolution
law~\eqref{11.14}, can be derived in the following way. Differentiating the
matrix form of~\eqref{11.14} with respect to $t$ and applying
equation~\eqref{5.3}, we obtain the
\emph{matrix-bundle Schr\"odinger equation for the density morphism}
as
	\begin{equation}	\label{11.15}
\ih\frac{d\mmor{\Rho}_\gamma(t)}{dt} =
[ \mbHam_\gamma(t) , \mmor{\Rho}_\gamma(t) ]_{\_}
	\end{equation}
with $\mbHam_\gamma(t)$ being the matrix-bundle Hamiltonian (given
by~\eqref{5.2}). This equation is the matrix-bundle analogue of~\eqref{11.5},
to which it is equivalent as it can be proved via the substitution of the
matrix form of~\eqref{11.11} into the one of~\eqref{11.5} (see
also~\eqref{6.2''}). Consequently, the results of Sect.~\ref{V} show
that~\eqref{11.14} gives the general solution ~\eqref{11.15} with respect to
$\Rho_\gamma(t)$.

	The matrix equation~\eqref{11.15} can be written into an invariant
form too. To this end we shall use the following result which is a simple
corollary of~\eqref{11.15}, \eqref{5.8}, and~\eqref{5.10}: If $\Psi$  is a
section along paths and one of the equations
 $\mor{D}_{t}^{\gamma}( \Psi_\gamma(t) ) = 0$,
 $\mor{D}_{t}^{\gamma}[ \Rho_\gamma(t) (\Psi_\gamma(t) ] = 0$,
or~\eqref{11.15} is valid, then the remaining two of them are equivalent.
From here follows that~\eqref{11.15} is equivalent to the system
	\begin{subequations}	\label{11.16}
\begin{align}	\label{11.16a}
(\mor{D}_{t}^{\gamma}\circ\Rho_\gamma(t)) \Psi_\gamma(t) &= 0, \\
		\label{11.16b}
\mor{D}_{t}^{\gamma}(\Psi_\gamma(t)) &= 0.
\end{align}
	\end{subequations}
(Note: $\gamma$ is not a summation index here and bellow!)

	If we denote by  $\tilde{\Rho}_\gamma(t)$ the restriction of
 ${\Rho}_\gamma(t)\colon\bHil_{\gamma(t)}\to\bHil_{\gamma(t)}$
on the set of (state) sections along $\gamma$ which are (linearly)
transported along $\gamma$ by means of the evolution transport, i.e the ones
satisfying~\eqref{11.16b}, then~\eqref{11.16} is equivalent to
	\begin{equation}	\label{11.17}
\tilde{\mor{D}}_t^\gamma( \tilde{\Rho}_\gamma(t) ) = 0
	\end{equation}
where $\tilde{\mor{D}}$ is the defined by~\eqref{6.3} differentiation along paths
of bundle morphisms (along paths in the present case). The above discussion
shows the equivalence of~\eqref{11.17} and~\eqref{11.15}, a fact which is also
an evident corollary of~\eqref{6.5} and~\eqref{5.10}. The
equation~\eqref{11.17} can be called a
\emph{(bundle) Schr\"odinger equation for the density morphism}.

	A simple verification proves that the linear map
 $\Rho_\gamma(t_0)\mapsto\Rho_\gamma(t)$, defined by~\eqref{11.14},
satisfies~\eqref{3.1} and~\eqref{3.2}. So, freely speaking, we may say that
this is a `transport-like' map by means of which $\Rho_\gamma$  is
`transported' along $\gamma$. This is something like `representation' of the
evolution transport in the space of morphisms along $\gamma$. The rigorous
study of this problem (see the end of subsection~\ref{VII.1}
and~\cite[sect.~3]{bp-TP-morphisms}) reviles that the pointed map is a
linear transport along $\gamma$ in the fibre bundle of bundle morphisms
(along paths) of $\HilB$. With respect to this transport along paths the
density morphism (along $\gamma$) is a linearly transported section (along
$\gamma$) of the bundle of morphisms (along $\gamma$) of $\HilB$.

\subsection{Representations in the different pictures of motion}
\label{XI.3}

	Let us turn now our attention to the description of mixed states in
the different pictures of motion (see Sect.~\ref{VII}).

	In the Schr\"odinger picture of the Hilbert bundle (resp.\ space)
description of quantum mechanics, which, in fact, was investigated until now
in this section, the motion of a quantum system is described by pairs like
$(\Rho_\gamma(t),\mor{A}_\gamma(t))$
(resp.\ $(\rho_\gamma(t),\ope{A}_\gamma(t))$)
of generally time-depending morphisms along paths of $\HilB$ (resp.\ operators
acting on $\Hil$) representing the density morphism (resp.\ operator) and some
dynamical variable $\dyn{A}$.

	The transition to the \emph{Heisenberg picture} is achieved via the
general formulae~\eqref{7.4} and~\eqref{7.7} for bundle morphisms along paths
in $\HilB$ and operators in $\Hil$, respectively. In particular, for the
density morphism $\Rho$ and operator $\rho$ they, respectively, give:
	\begin{align}
		\label{11.18}
\Rho_{\gamma,t}^{\mathrm{H}}(t_0) &:=
\mor{U}_{\gamma}^{-1}(t,t_0)\circ\Rho_\gamma(t)\circ\mor{U}_\gamma(t,t_0) =
\Rho_\gamma(t_0)
\colon\bHil_{\gamma(t_0)}\to\bHil_{\gamma(t_0)},	\\
		\label{11.19}
\rho_{t}^{\mathrm{H}}(t_0) &:=
\ope{U}^{-1}(t,t_0)\circ\rho(t)\circ\ope{U}(t,t_0) =
\rho_\gamma(t_0)
\colon\Hil\to\Hil
	\end{align}
where~\eqref{11.14} and~\eqref{11.6} were used. Consequently in the
Heisenberg picture the Hilbert bundle and Hilbert space descriptions
are by means of pairs like
\(
\left(
\Rho_{\gamma,t}^\mathrm{H}(t_0),\mor{A}_{\gamma,t}^\mathrm{H}(t_0)
\right)
=
\left(
\Rho_{\gamma}(t_0),\mor{A}_{\gamma,t}^\mathrm{H}(t_0)
\right)
\)
and
\(
\left( \rho_{t}^\mathrm{H}(t_0),\ope{A}_{t}^\mathrm{H}(t_0) \right) =
\left( \rho(t_0),\ope{A}_{t}^\mathrm{H}(t_0) \right),
\)
respectively,
in which the time dependence is entirely shifted from the density morphisms
and operators to the observables. So, in this picture the density
morphisms and operators are constant(s of motion), do not evolve in time,
while the observer's evolution is governed by the Heisenberg equation of
motion for them (see~\eqref{7.10}, or~\eqref{7.8}, or~\eqref{7.13b9}).

	Of course, in the Heisenberg picture the mean values remain
unchanged:
	\begin{equation}	\label{11.20}
\langle\mor{A}_{\gamma,t}^{\mathrm{H}}(t_0)\rangle
_{ \Rho_{\gamma,t}^{\mathrm{H}} }^{t_0} =
\langle\ope{A}_{t}^{\mathrm{H}}(t_0)\rangle
_{ \rho_{t}^{\mathrm{H}} }^{t_0} =
\langle\mor{A}_{\gamma}^{}(t)\rangle_{ \Rho_{\gamma}^{} }^{t} =
\langle\ope{A}(t)\rangle_{ \rho }^{t},
	\end{equation}
where
	\begin{equation}	\label{11.21}
\langle\mor{A}_{\gamma,t}^{\mathrm{H}}(t_0)\rangle
_{ \Rho_{\gamma,t}^{\mathrm{H}} }^{t_0} :=
\Tr\left( \Rho_\gamma(t_0)\circ\mor{A}_{\gamma,t}^{\mathrm{H}}(t_0) \right),
\quad
\langle\ope{A}_{t}^{\mathrm{H}}(t_0)\rangle
_{ \rho_{t}^{\mathrm{H}} }^{t_0} :=
\Tr\left( \rho(t_0)\circ\ope{A}_{t}^{\mathrm{H}}(t_0) \right).
	\end{equation}
The chain equation~\eqref{11.20} is a corollary of the invariance of the
trace of a product (composition) of operators with respect to a cyclic
permutation of the multipliers.

	The shift from the Schr\"odinger to `general' picture is done by the
general equations~\eqref{7.16} and~\eqref{7.19}. Hence, in the
$\mor{V}$-picture of motion the density morphism $\Rho$ and operator $\rho$,
respectively, are
	\begin{align}
	\label{11.22}
\Rho_{\gamma,t}^\mor{V}(t_1) &=
\mor{V}_\gamma(t_1,t) \circ \Rho_\gamma(t) \circ
\mor{V}_\gamma^{-1}(t_1,t)\colon
\bHil_{\gamma(t_1)} \to \bHil_{\gamma(t_1)}, \\
	\label{11.23}
\rho_{t}^\ope{V}(t_1) &=
\ope{V}_\gamma(t_1,t) \circ \rho_\gamma(t) \circ \ope{V}^{-1}(t_1,t)
\colon \Hil \to \Hil.
	\end{align}

	Since in the $\mor{V}$-picture the Hilbert bundle (resp.\ space)
description is via pairs like
\(
\left( \Rho_{\gamma,t}^\mor{V}(t_1) , \mor{A}_{\gamma,t}^\mor{V}(t_1) \right)
\)
\(\left(
\text{resp.\
\( \left( \rho_{t}^\ope{V}(t_1) , \ope{A}_{t}^\ope{V}(t_1) \right) \)%
}
\right),
\)
the mean values of the observables remain unchanged, as in the Heisenberg
picture:
	\begin{equation}	\label{11.24}
\langle \mor{A}_{\gamma,t}^\mor{V}(t_1) \rangle
_{\Rho_{\gamma,t}^\mor{V}}^{t_1} =
\langle \ope{A}_{t}^\ope{V}(t_1) \rangle _{\rho_{t}^\ope{V}}^{t_1} =
\langle \mor{A}_{\gamma}(t) \rangle
_{\Rho_{\gamma}}^{t} =
\langle \ope{A}(t) \rangle _{\rho}^{t},
	\end{equation}
where
	\begin{equation}	\label{11.25}
	\begin{split}
\langle \mor{A}_{\gamma,t}^\mor{V}(t_1) \rangle
_{\Rho_{\gamma,t}^\mor{V}}^{t_1} &:=
\Tr\left(
\Rho_{\gamma,t}^{\mor{V}}(t_1) \circ \mor{A}_{\gamma,t}^\mor{V}(t_1)
\right),
\\
\langle \ope{A}_{t}^\ope{V}(t_1) \rangle _{\rho_{t}^\ope{V}}^{t_1} &:=
\Tr\left(
\rho_{t}^\ope{V}(t_1) \circ \ope{A}_{t}^\ope{V}(t_1)
\right).
	\end{split}
	\end{equation}

	In the $\mor{V}$-picture the density morphisms,
operators, and observables generally change in time. For all of them this
change is governed by the equations~\eqref{7.29} and~\eqref{7.27}, but for
the density morphisms and operators they can be written in a more concrete
form. For this purpose we have to calculate the last terms in the r.h.s.
of~\eqref{7.29} and~\eqref{7.27}.

	Using~\eqref{7.16} and~\eqref{11.5}, we obtain
\[
\left( \frac{\partial\rho(t)}{\partial t} \right)_{\!t}^{\!\ope{V}} (t_1) =
\ope{V}(t_1,t)\circ
\frac{\partial\rho(t)}{\partial t}\circ\ope{V}^{-1}(t_1,t) =
\iih
\left[ \Ham_{t}^{\ope{V}}(t_1) , \rho_{t}^{\ope{V}}(t_1) \right]_{\_}.
\]
Analogously, applying~\eqref{7.22} and the just get equation, we find
\[
\left( \frac{\partial\rho(t)}{\partial t} \right)
_{\!\gamma,t}^{\!\mor{V}} (t_1) =
l_{\gamma(t_1)}^{-1} \circ
\left( \frac{\partial\rho(t)}{\partial t} \right)_{\!t}^{\!\ope{V}} (t_1)
\circ l_{\gamma(t_1)} =
\iih\left[
\bHam_{\gamma,t}^{\mor{V}}(t_1) , \Rho_{\gamma,t}^{\mor{V}}(t_1)
\right]_{\_}.
\]
At last, substituting the above two equations into~\eqref{7.29}
and~\eqref{7.27}, we, respectively, get
	\begin{align}
		\label{11.26}
\ih
\frac{\partial\Rho_{\gamma,t}^{\mor{V}}(t_1)}{\partial t} &=
\left[
\widetilde{\bHam}_{\gamma,t}^{\mor{V}}(t_1) ,
\Rho_{\gamma,t}^{\mor{V}}(t_1)
\right]_{\_},	\\
		\label{11.27}
\ih
\frac{\partial\rho_{t}^{\ope{V}}(t_1)}{\partial t} &=
\left[
\widetilde{\Ham}_{t}^{\ope{V}}(t_1) ,
\rho_{t}^{\ope{V}}(t_1)
\right]_{\_}
	\end{align}
where~\eqref{7.25} and~\eqref{7.28-29} were taken into account. These are the
equations of motion for the density morphism and operator in the $V$-picture.

	If the evolution transport and operator are known (in the
$V$-picture), then combining, from one hand,~\eqref{11.22}, \eqref{11.14}
and~\eqref{7.35} and, from the other hand, \eqref{11.23}, \eqref{11.6},
and~\eqref{7.34}, we get the general solution of~\eqref{11.26}
and~\eqref{11.27}, respectively, in the form
	\begin{align}
		\label{11.28}
\Rho_{\gamma,t}^{\mor{V}}(t_1) &=
\mor{U}_\gamma^\mor{V}(t,t_1,t_0)\circ
\Rho_{\gamma,t_0}(t_1)\circ
\left(\mor{U}_\gamma^\mor{V}(t,t_1,t_0)\right)^{-1},	\\
		\label{11.29}
\rho_{t}^{\ope{V}}(t_1) &=
\ope{U}^\ope{V}(t,t_1,t_0)\circ
\rho_{t_0}(t_1)\circ
\left(\ope{U}^\ope{V}(t,t_1,t_0)\right)^{-1}.
	\end{align}

	As one can expect, in the case of Heisenberg picture, due
to~\eqref{7.35a}, these formulae reproduce~\eqref{11.18} and~\eqref{11.19}
respectively.

\section{Curvature of the evolution transport}
\label{IX}
\setcounter{equation} {0}

	Let $\eta\colon J\times J^\prime\to \base$ with $J$ and $J^\prime$
 being $\mathbb{R}$-intervals. According to~\cite[Sect.~3]{bp-LTP-Cur+Tor} the
curvature of the (bundle) evolution transport $\mor{U}$ is
$R\colon \eta\mapsto R^\eta$, with
$R^\eta\colon (s,t)\mapsto R^\eta(s,t),\ (s,t)\in J\times J^\prime$,
where
	\begin{equation}	\label{9.1}
R^\eta(s,t) :=
\mor{D}_s^{\eta(\cdot,t)} \!\circ \mor{D}_t^{\eta(s,\cdot)} -
\mor{D}_t^{\eta(s,\cdot)} \!\circ \mor{D}_s^{\eta(\cdot,t)}
\colon \Sec^2\HilB\to 
\pi^{-1}(\eta(s,t))
	\end{equation}
Here $\mor{D}$ is the differentiation along paths assigned to $\mor{U}$
by~(\ref{5.7}) .

	In a local field of bases the local components of the curvature
are~\cite[equation~3.3]{bp-LTP-Cur+Tor}
	\begin{multline}
\left( R^\eta(s,t) \right)_{{\ }b}^a  =
\frac{\partial}{\partial s}
\left[ \Gamma_{{\ }b}^a(t;\eta(s,\cdot)) \right] -
\frac{\partial}{\partial t}
\left[ \Gamma_{{\ }b}^a(s;\eta(\cdot,t)) \right]
\\	\label{9.2}
+
\Gamma_{{\ }c}^a(s;\eta(\cdot,t)) \Gamma_{{\ }b}^c(t;\eta(s,\cdot)) -
\Gamma_{{\ }c}^a(t;\eta(s,\cdot)) \Gamma_{{\ }b}^c(s;\eta(\cdot,t)).
	\end{multline}

	Physically we interpret
$\eta(s,\cdot)\colon t\mapsto \eta(s,t)$ and
$\eta(\cdot,t)\colon s\mapsto \eta(s,t)$
as world lines (trajectories) of observers with proper times $t$ and $s$
respectively. So, using the fundamental relation~(\ref{5.10}) we can
explicitly calculate the curvature. The easiest way to do this is to
choose the bases $\{e_a(x)\}$ and $\{f_a(t)\}$ such that
$\boldsymbol{l}_x(t)=[\delta_b^a]=\openone$ (see remark~\ref{rem5.1}).
Then
$\boldsymbol{E}(t)=0$ and
\(
\mbHam_{\gamma}(t) =
\bHamM(t)=\HamM(t).
\)
Hence, now~(\ref{9.2}) reduces to
	\begin{equation}	\label{9.3}
\boldsymbol{R}^\eta(s,t)=\frac{1}{(-\ih)^2}
[ \HamM(s) , \HamM(t) ]_{\_} ,
	\end{equation}
where we have assumed, as usual, that $\HamM(s)$ is independent of the
observers trajectory $\gamma$. (This equality is valid only in the special
basis in which it is derived!)

	From here it follows that the
\emph{%
evolution transport is curvature free if and only if the values of the
Hamiltonian operator at different moments commute,%
}
viz.\
	\begin{equation}	\label{9.4}
{R}^\eta = 0 \iff [ {\Ham}(s) , {\Ham}(t) ]_{\_} = 0.
	\end{equation}

	In particular this is true for time-independent Hamiltonians, i.e.\
for $\partial\Ham(t)/\partial t = 0$ . According to~(\ref{6.085}) the bundle
formulation of~\eqref{9.4} is
	\begin{equation}	\label{9.5}
{R}^\eta = 0 \iff
\left[
{\Breve{\bHam}}_{\gamma,s}(r) , {\Breve{\bHam}}_{\gamma,t}(r)
\right]_{\_} = 0.
	\end{equation}
for some (and hence any) path $\gamma\colon J\to\base$. Here $r,s,t\in J$ and
${\Breve{\bHam}}_{\gamma,t}(r)$ is (the transported by means of
$l_{s\to t}^{\gamma}$ Hamiltonian which is) calculated via~(\ref{6.082}).

	Consider now a curvature free evolution transport on
$W=\eta(J,J')$, \ie  $R^\eta(s,t)\equiv0$ for every $(s,t)\in J\times J'$.
From~\cite[proposition~3.3]{bp-LTP-Cur+Tor} we know that in this case
there exists a field of base $\{e_i\}$ over $W$ in which the transport's
coefficients vanish along any path $\gamma$ in $W$. In it, due to~\eref{5.3}
and~\eref{5.10}, we have
	\begin{equation}	\label{9.6}
\widetilde{\mmor{U}_\gamma}(t,t_0) = \openone,	\quad
\widetilde{\boldsymbol{\Gamma}}_\gamma(t) = \boldsymbol{0},	\quad
\widetilde{\mbHam_\gamma}(t) = \boldsymbol{0}
\qquad\text{for every $\gamma$}.
	\end{equation}
Notice, equation~\eref{7.3} is valid for \emph{arbitrary} evolution
transports along any \emph{fixed} path in appropriate bases along it,
while~\eref{9.6} holds only for \emph{curvature free} evolution transports in
a suitably chosen field of bases in a \emph{whole} set $W$. The connection of
the special bases in which~\eref{9.6} is true with the Heisenberg picture is
the same as discussed in Subsect~\ref{VII.2} for the bases in
which~\eref{7.3} are satisfied.

	We want to emphasize on the fact that according to~\eref{5.2} the
local vanishment of the matrix\ndash bundle Hamiltonian does not imply the
same property for the Hamiltonian as an operator or morphism.

\section {Conclusion}
\label{conclusion-IV}
\setcounter{equation} {0}

	In the present work we have seen that the bundle formulation of
quantum mechanics admits natural description of mixed states. We have
derived different versions of the equation of motion for the density morphism
along paths, which now replaces the conventional density operator. The mixed
states were considered from the view-point of different bundle pictures of
motion. Here we have calculated the curvature of the evolution transport. It
turns to be curvature free iff the values of the Hamiltonian at different
moments commute.

	This paper ends the introduction of general formalism of Hilbert
bundle description of nonrelativistic quantum mechanics. The interpretation of
this description and its possible further developments will be given
elsewhere.

\section*{Acknowledgments}

	This work was partially supported by the National Foundation for
Scientific Research of Bulgaria under Grant No.~F642.


\addcontentsline{toc}{section}{References}

\bibliography{bozhopub,bozhoref}

\begin{thebibliography}{10}

\bibitem{bp-BQM-introduction+transport}
Bozhidar~Z. Iliev.
\newblock Fibre bundle formulation of nonrelativistic quantum mechanics. {I}.
  {Introduction}. {The} evolution transport.
\newblock (LANL xxx archive server, E-print No. quant-ph/9803084) Submitted to
  J. Physics A: Math. \& Gen., March 1998.

\bibitem{bp-BQM-equations+observables}
Bozhidar~Z. Iliev.
\newblock Fibre bundle formulation of nonrelativistic quantum mechanics. {II}.
  {Equations} of motion and observables.
\newblock (LANL xxx archive server, E-print No. quant-ph/9804062) Submitted to
  J. Physics A: Math. \& Gen., April 1998.

\bibitem{bp-BQM-pictures+integrals}
Bozhidar~Z. Iliev.
\newblock Fibre bundle formulation of nonrelativistic quantum mechanics. {III}.
  {Pictures} and integrals of motion.
\newblock (LANL xxx archive server, E-print No. quant-ph/9806046) Submitted to
  J. Physics A: Math. \& Gen., October 1998.

\bibitem{Messiah-QM}
A.~M.~L. Messiah.
\newblock {\em Quantum mechanics}.
\newblock Interscience, New York, 1958.
\newblock (Vol.~I and vol.~II.).

\bibitem{Dirac-PQM}
P.~A.~M. Dirac.
\newblock {\em The principles of quantum mechanics}.
\newblock Oxford at the Clarendon Press, Oxford, fourth edition, 1958.

\bibitem{Prugovecki-QMinHS}
E.~Prugove\v{c}ki.
\newblock {\em Quantum mechanics in Hilbert space}, volume~92 of {\em Pure and
  applied mathematics}.
\newblock Academic Press, New York-London, second edition, 1981.

\bibitem{Uhlmann-91a}
A.~Uhlmann.
\newblock A gauge field governing parallel transport along mixed states.
\newblock {\em Lett. Math. Phys.}, 21(3):229--236, March 1991.

\bibitem{Anandan-90a}
J.~Anandan.
\newblock A geometric view of quantum mechanics.
\newblock In Anandan~J. S., editor, {\em Quantum coherence}, Proceedings of the
  Conference on Fundamental Aspects of Quantum theory, Columbia, South
  Carolina, 14--16~December 1989. World Scientific, Singapore, 1990.
\newblock (See also Preprint MPI-PAE/PTh 77/90, 1990 and Found. Phys.,
  21(11):1265--1284, 1991).

\bibitem{bp-TP-morphisms}
Bozhidar~Z. Iliev.
\newblock Transports along paths in fibre bundles. {III}.~{Consistency} with
  bundle morphisms.
\newblock JINR Communication E5-94-41, Dubna, 1994.
\newblock (LANL xxx archive server, E-print No. dg-ga/9704004).

\bibitem{bp-LTP-Cur+Tor}
Bozhidar~Z. Iliev.
\newblock Linear transports along paths in vector bundles. {III}.~{Curvature}
  and torsion.
\newblock JINR Communication E5-93-261, Dubna, 1993.

\end{thebibliography}
\bibliographystyle{unsrt}

\addcontentsline{toc}{subsubsection}{\vspace{1ex}This article ends at page}

\end{document}